\documentclass[pre,twocolumn,showpacs,superscriptaddress,preprintnumbers,floatfix]{revtex4}

\usepackage{dcolumn}
\usepackage{url}
\usepackage{amsmath}
\usepackage{amssymb}
\usepackage{graphicx}
\usepackage{bm}   
\usepackage{bbm}   
\usepackage{verbatim}
\usepackage{stmaryrd}
\usepackage{amsthm}

\theoremstyle{plain}    
\theoremstyle{plain}    
\theoremstyle{plain} 	
\theoremstyle{plain} 	
\theoremstyle{plain} 	
\theoremstyle{plain} 	
\theoremstyle{plain} 	
\theoremstyle{plain} 	
\theoremstyle{plain} 	
\theoremstyle{plain}	
\theoremstyle{plain}	 
\theoremstyle{plain}	

\newcommand{\MeasSymbol}   { {X} }
\newcommand{\meassymbol}   { {x} }

\newcommand{\Past}	{ \stackrel{\leftarrow} {\MeasSymbol} }
\newcommand{\past}	{ {\stackrel{\leftarrow} {\meassymbol}} }

\newcommand{\Future}	{ \stackrel{\rightarrow}{\MeasSymbol} }
\newcommand{\future}	{ \stackrel{\rightarrow}{\meassymbol} }

\newcommand{\AllPasts}	{ { \stackrel{\leftarrow} {\rm {\bf \MeasSymbol}} } }

\newcommand{\CausalState}	{ \mathcal{S} }

\newcommand{\causalstate}	{ \sigma }
\newcommand{\CausalStateSet}	{ \boldsymbol{\CausalState} }
\newcommand{\AlternateState}	{ {\cal R} }
\newcommand{\alternatestate}	{ \rho }
\newcommand{\AlternateStateSet}	{ \boldsymbol{\AlternateState} }
\newcommand{\PrescientState}	{ \widehat{\AlternateState} }

\newcommand{\PrescientStateSet}	{ \boldsymbol{\PrescientState}}

\newcommand{\Prob}		{ {\rm P}}

\newcommand{\Cmu}		{ {C_\mu}}
\newcommand{\hmu}		{ {h_\mu}}
\newcommand{\EE}		{ {\bf E}}

\newcommand{\InfoGain}[2] { \mathcal{D} \left( {#1} || {#2} \right) }

\newcommand{\Partition}	{ \AlternateState }
\newcommand{\partitionstate}	{ \alternatestate }

\begin{document}

\title{Structure or Noise?}

\author{Susanne Still}
\email{sstill@hawaii.edu}
\affiliation{Information \& Computer Sciences,
University of Hawaii at Manoa, Honolulu, HI 96822}

\author{James P. Crutchfield}
\email{chaos@cse.ucdavis.edu}
\affiliation{Complexity Sciences Center \& Physics Department,
University of California Davis, One Shields Avenue, Davis, CA 95616}

\date{\today}

\bibliographystyle{unsrt}

\begin{abstract}
We show how rate-distortion theory provides a mechanism for automated theory
building by naturally distinguishing between regularity and randomness. We
start from the simple principle that model variables should, as much as
possible, render the future and past conditionally independent. From this,
we construct an objective function for model making whose extrema embody the
trade-off between a model's structural complexity and its predictive power.
The solutions correspond to a hierarchy of models that, at each level of
complexity, achieve optimal predictive power at minimal cost. In the limit
of maximal prediction the resulting optimal model identifies a process's
intrinsic organization by extracting the underlying causal states. In this
limit, the model's complexity is given by the statistical complexity, which
is known to be minimal for achieving maximum prediction. Examples show how
theory building can profit from analyzing a process's
\emph{causal compressibility}, which is reflected in the optimal models'
rate-distortion curve---the process's characteristic for optimally balancing
structure and noise at different levels of representation.
\end{abstract}

\preprint{Santa Fe Institute Working Paper 07-08-020}
\preprint{arxiv.org:0708.0654 [physics.gen-ph]}


\maketitle





\section{Introduction}

Progress in science is often driven by the discovery of novel patterns. 
Historically, physics has relied on the creative mind of the theorist to 
articulate mathematical models that capture nature's regularities in physical 
principles and laws. But the last decade has witnessed a new era in collecting 
truly vast data sets. Examples include contemporary experiments in particle 
physics \cite{LHC07a} and astronomy \cite{LSST07a}, but range to genomics,
automated language translation \cite{NIST06a}, and web social organization
\cite{Newm01a}. In all these, the volume of data far exceeds 
what any human can analyze directly by hand.

This presents a new challenge---automated pattern discovery and model building.
A principled understanding of model making is critical to provide theoretical
guidance for developing automated procedures. In this Letter, we show how basic 
information-theoretic optimality criteria provide a method for automatically
constructing a hierarchy of models that achieve different degrees of
abstraction. Importantly, we show that in appropriate limits the method
recovers a process's causal organization. Without this connection, it would be
only another approach to statistical inference, with its own ad hoc assumptions
about the character of natural pattern.

Our starting point is the observation that natural systems store, process,
and produce information---they compute intrinsically
\cite{Crut88a,Crut92c,Crut98d}. Theory building, then, faces
the challenge of extracting from that information the structures
underling its generation. Any physical theory delineates mechanism from 
randomness by identifying what part of an observed phenomenon is due to 
the underlying process's structure and what is irrelevant. Irrelevant parts 
are considered noise and typically modeled probabilistically. Successful 
theory building therefore depends centrally on deciding what is structure
and what is noise; often, an implicit distinction.

What constitutes a good theory, though? Which information is relevant? 
One can answer this question for time series prediction: Information about 
the future of the time series is relevant. Beyond forecasting, though,
models are often put to the test by assessing how well they predict new
data and, hence, it is of general importance that a model capture information
which aids prediction. Typically, there are many models that explain a given
data set, and between two models that are equally predictive, one favors the
simpler, smaller, less structurally complex model \cite{Occam,Domi99a}.
However, a more complex model can achieve smaller prediction error than a
less complex model. The trade-off between model complexity and prediction
error is tantamount to finding a distinction between causal structure and noise.

The trade-off between assigning a causal mechanism to the occurrence of an
event or explaining the event as being merely random has a long history, but 
how one implements the trade-off is still a very active topic. Nonlinear time
series analysis \cite{Casd91a,Spro03a,Kant06a}, to take one example, attempts to
account for long-range correlations produced by nonlinear dynamical
systems---correlations not adequately modeled by assumptions such as linearity
and independent, identically distributed (i.i.d.) data. Success in this
endeavor requires directly addressing the notion of structure and pattern
\cite{Casd91a,Crut87a}.

Examination of the essential goals of prediction led to a principled definition
of structure that captures a dynamical system's causal organization in part by
discovering the underlying \emph{causal states} \cite{Crut88a,Crut92c,Crut98d}.
In \emph{computational mechanics} a process $\Prob(\Past,\Future)$ is viewed
as a communication channel \cite{Cove06a,Crut98d}: it transmits information
from the
\emph{past} $\Past = \ldots \MeasSymbol_{-3} \MeasSymbol_{-2} \MeasSymbol_{-1}$
to the \emph{future}
$\Future = \MeasSymbol_0 \MeasSymbol_1 \MeasSymbol_2 \ldots$ by storing it in
the present. For the purpose of forecasting the future two different pasts,
say $\past$ and $\past^\prime$, are equivalent if they result in the same
prediction \cite{Crut88a}. In general this prediction is probabilistic, given
by the conditional future distribution $\Prob(\Future|\past)$. The resulting
equivalence relation $\past \sim \past^\prime$ groups all histories that give
rise to the same conditional future distribution:
\begin{equation}
\epsilon(\past) = \{ \past^\prime: \Pr(\Future|\past) = \Pr(\Future|\past^\prime) \}.
\label{eps}
\end{equation}   
The resulting partition of the space $\AllPasts$ of pasts defines the 
process's \emph{causal states} $\CausalStateSet = \Prob(\Past,\Future) / \sim$.
 
The causal states constitute a model that is maximally predictive by means of
capturing all the information that the past of a time series contains about
the future. As a result, knowing the causal state renders past and future
conditionally independent, a property we call \emph{causal shielding}, because
the causal states have the Markovian property that they \emph{shield} past and
future \cite{Crut98d}:
\begin{equation}
\Prob(\Past,\Future|\CausalState) = \Prob(\Past|\CausalState) \Prob(\Future|\CausalState),
\label{shield}
\end{equation}
where $\CausalState \in \CausalStateSet$. This is related to the fact that the
causal-state partition is optimally predictive. To see this, note that
Eq. (\ref{shield}) implies
$\Prob(\Future|\Past,\CausalState) = \Prob(\Future|\CausalState)$. Furthermore,
note that, by definition, for \emph{any} partition $\AlternateStateSet$ of
$\AllPasts$ with states $\AlternateState$, when the past is known, then the
future distribution is not altered by the history-space partitioning: 
\begin{equation}
\Prob(\Future|\Past, \AlternateState) = \Prob(\Future|\Past) ~.
\label{markov}
\end{equation} 
This implies for the causal states that
$\Prob(\Future|\Past, \CausalState) = \Prob(\Future|\Past)$ and thus
$\Prob(\Future|\CausalState) = \Prob(\Future|\Past)$. Therefore, causal
shielding is equivalent to the fact \cite{Crut98d} that the causal states
capture \emph{all} of the information $I[\Past;\Future]$ that is shared
between past and future: $I[\CausalState;\Future] = I[\Past;\Future]$, 
the process's \emph{excess entropy} $\EE$ or \emph{predictive information} 
\cite[and references therein]{Crut01a,BialekNemTishby2001}.

The causal states are {\em unique and minimal sufficient statistics} for
time series prediction, capturing all of a process's predictive information
at maximum efficiency \cite{Crut98d}. The causal-state partition has the
smallest \emph{statistical complexity},
$\Cmu := H(\CausalState) \leq H[\PrescientState]$, compared to all other 
equally predictive partitions $\PrescientStateSet$. $\Cmu$ measures the
minimal amount of information that must be stored in order to communicate all
of the excess entropy from the past to the future. Briefly stated, the causal
states serve as the basis against which alternative models should be compared.

\section{Constructing causal models using rate-distortion theory}

There are many scenarios in which one does not need to or explicitly
does not want to capture \emph{all} of the predictive information. How can 
we approximate the causal states in a controlled way? 

In this Letter, we show how to systematically construct smaller models, which
are necessarily less predictive, but which are optimal in the sense that they
capture, at a fixed model complexity, the maximum possible amount of predictive
information. Importantly, in the limit that removes the constraint on model
complexity, our method retrieves the exact causal-state partition. 

Appealing to information theory again, we frame this in terms of communicating
a model over a channel with limited capacity. Rate-distortion theory
\cite{Shannon48} provides a principled
way to find a lossy compression of an information source such that the
resulting code is minimal at fixed fidelity to the original signal.

The compressed representation, denote it $\AlternateStateSet$, is in general
specified by a \emph{probabilistic} map $\Prob(\AlternateState|\past)$ from the
input message, here the past $\past$, to code words, here the model's states
$\AlternateState$ with values $\alternatestate \in \AlternateStateSet$. 
In contrast, Eq. (\ref{eps}) specifies models that are described by a
deterministic map from histories to states: The causal states 
$\causalstate \in \CausalStateSet$ induce a deterministic partition of
$\AllPasts$ \cite{Crut98d}, as one can show that
$\Prob(\causalstate|\past) = \delta_{\causalstate,\epsilon(\past)}$.
The mapping $\Prob(\AlternateState|\past)$ specifies a model, and the 
\emph{coding rate} $I[\Past;\AlternateState]$ measures its complexity, 
which in turn is related to its statistical complexity via 
$I[\Past;\AlternateState] = H[\AlternateState] - H[\AlternateState|\Past] =  
\Cmu(\AlternateState) - H[\AlternateState|\Past]$.
For deterministic partitions the statistical complexity and the coding rate
are equal, because then $H[\AlternateState|\Past] = 0$. However, for more
general, nondeterministic partitions, $H[\AlternateState|\Past] \neq 0$,
meaning that the probabilistic nature of the mapping curtails some of the
model's complexity, and the coding rate $I[\Past;\AlternateState]$ captures
this.

To illustrate this point, consider the extreme of uniform assignments:
$\Prob(\AlternateState|\past) = 1/c$, for any given $\past$, where 
$c = |\AlternateStateSet|$. In this case, even if there are many
states---large statistical complexity $H[\AlternateState] = \log_2(c)$---they
are indistinguishable:
\mbox{$\Prob(\future| \AlternateState) = \langle
\Prob(\future|\past) \rangle_{\Prob(\past)}$}, for all $\AlternateState$,
due to the large uncertainty about the state, given the past. This is reflected
in $H[\AlternateState|\Past]  = \log_2(c)$. In effect, the model has only one
state (the average $\langle \Prob(\future|\past) \rangle_{\Prob(\past)}$)
and its statistical complexity vanishes, which is reflected in the coding
rate: $I[\Past;\AlternateState] = 0$.

Rate-distortion theory allows us to back away from the best (causal-state)
representation toward less complex models by controlling the coding rate:
Simpler models are distinguished from more complex ones by the fact that they 
can be transmitted more concisely. However, less complex models are also 
associated with a larger error. Rate-distortion theory quantifies the loss 
by a \emph{distortion function} $d(\past;\alternatestate)$. The
coding rate is then minimized \cite{Cove06a} over the assignments
$\Prob(\Partition|\Past)$ at fixed average distortion 
$D[\Past; \Partition] = \left\langle d(\past;\alternatestate) \right\rangle_{\Prob(\past,\alternatestate)}$.

In building predictive models, the loss should be measured by how much the
resulting models deviate from accurate prediction. We take the shielding
property, Eq. (\ref{shield}), of the causal-state partition as the goal for
any predictive model. This condition is equivalent to the statement that the
excess entropy \emph{conditioned on the model states} $\AlternateState$:
\begin{equation} 
I[\Past;\Future \! | \Partition]
  \!= \!\left\langle \!\!\! \left\langle 
  \log \left[ \frac{\Prob(\past,\future\! |\alternatestate)}
  {\Prob(\future\! | \alternatestate) \Prob(\past\! | \alternatestate)} \right] 
  \right\rangle_{\!\!\! \Prob(\future|\past)}
  \right\rangle_{\!\!\! \Prob(\past,\partitionstate)}
\label{ipredcond}
\end{equation}
vanishes for the causal-state partition:
$I[\Past;\Future \! | \CausalState] = 0$. This gives us our distortion measure: 
\begin{equation} 
d(\past; \alternatestate)  = \left\langle
   \log \left[ \frac{\Prob(\past,\future|\alternatestate)}
   { \Prob(\past|\alternatestate) \Prob(\future| \alternatestate) } \right]  
   \right\rangle_{\Prob(\future|\past)}
   .
\end{equation} 
From Eq. (\ref{markov}) this is the same as the relative entropy between the
conditional future distributions given the past and those given the model
states $\alternatestate$:
\begin{equation} 
\InfoGain{\Prob(\future|\past)}{\Prob(\future|\alternatestate)} = \left\langle
   \log \left[ \frac{\Prob(\future|\past)}
   { \Prob(\future| \alternatestate) } \right]  
   \right\rangle_{\Prob(\future|\past)}
   \label{dist-fkt}
\end{equation} 
Altogether, we solve the constrained optimization problem:
\vspace{-0.05in}
\begin{equation} 
\min_{\Prob(\Partition|\Past)}
  \left( I[\Past; \Partition] + \beta I[\Past; \Future|\Partition] 
  \right) ~,
\label{opt}
\end{equation}
where the Lagrange multiplier $\beta$ controls the trade-off between model
complexity and prediction error; i.e., the balance between structure and noise.

The conditional excess entropy of Eq. (\ref{ipredcond}) is the difference
between the process's excess entropy and the information
$I[\Partition;\Future]$ that the model states contain about the future:
$I [\Past ;\Future| \Partition] = I[\Past;\Future] - I[\Partition;\Future]$,
due to Eq. (\ref{markov}). The excess entropy $I[\Past;\Future]$ is a
property intrinsic to the process, however, and so not dependent on the model.
Therefore, the optimization problem in Eq. (\ref{opt}) 
is equivalent to maximizing the information that the model states carry about 
the future while minimizing information kept about the past. This maps
directly onto the \emph{information bottleneck} (IB) method \cite{IBN}---here
the future data is IB's ``relevant'' quantity with respect to which the
past is summarized.

In any case, the solution to the optimization principle is given by
(cf. \cite{IBN}):
\vspace{-0.10in}
\begin{equation} 
\Prob_\mathrm{opt}(\alternatestate|\past)
  = \frac{\Prob(\alternatestate)}{Z(\past,\beta)}
  e^{ -\beta E(\alternatestate,\past)} ~,
\label{OCF_States}
\end{equation}
where
\vspace{-0.10in}
\begin{align}
E(\alternatestate,\past) & =
  \InfoGain{\Prob(\Future|\past)}{\Prob(\Future|\alternatestate)} ~,
\label{OCF_Energy} \\
  \Prob(\Future|\alternatestate) & = \frac{1}{\Prob(\alternatestate)}
 \sum_{\past \in \AllPasts} \Prob(\Future|\past)
 \Prob(\alternatestate|\past) \Prob(\past) ~,~\mathrm{and} \\
 \Prob(\alternatestate) & = \sum_{\past \in \AllPasts}
 \Prob(\alternatestate|\past) \Prob(\past) ~.
\label{OCF_States_2}
\end{align}
Eqs. (\ref{OCF_States})-(\ref{OCF_States_2}) must be solved self-consistently,
and this can be done numerically \cite{IBN}.

Eq. (\ref{OCF_States}) specifies a family of models parametrized by $\beta$
with the form of Gibbs distributions. Within an analogy to statistical
mechanics \cite{Rose98}, $\beta$ corresponds to the inverse temperature, 
$E$ is the energy, and 
$Z = \left\langle e^{ -\beta E(\alternatestate,\past)}
  \right\rangle_{\Prob(\alternatestate)} $ 
the partition function. 
Finally, note that for linear Gaussian-distributed random variables the
optimal linear map can be computed analytically \cite{Chec05a}. These results
can be carried over to the temporal setting that concerns us here for linear
Gaussian processes following a rate-distortion approach similar to the
above \cite{Creu08a}.

\begin{figure*}[th]
\begin{center}
\resizebox{!}{2.20in}{\includegraphics{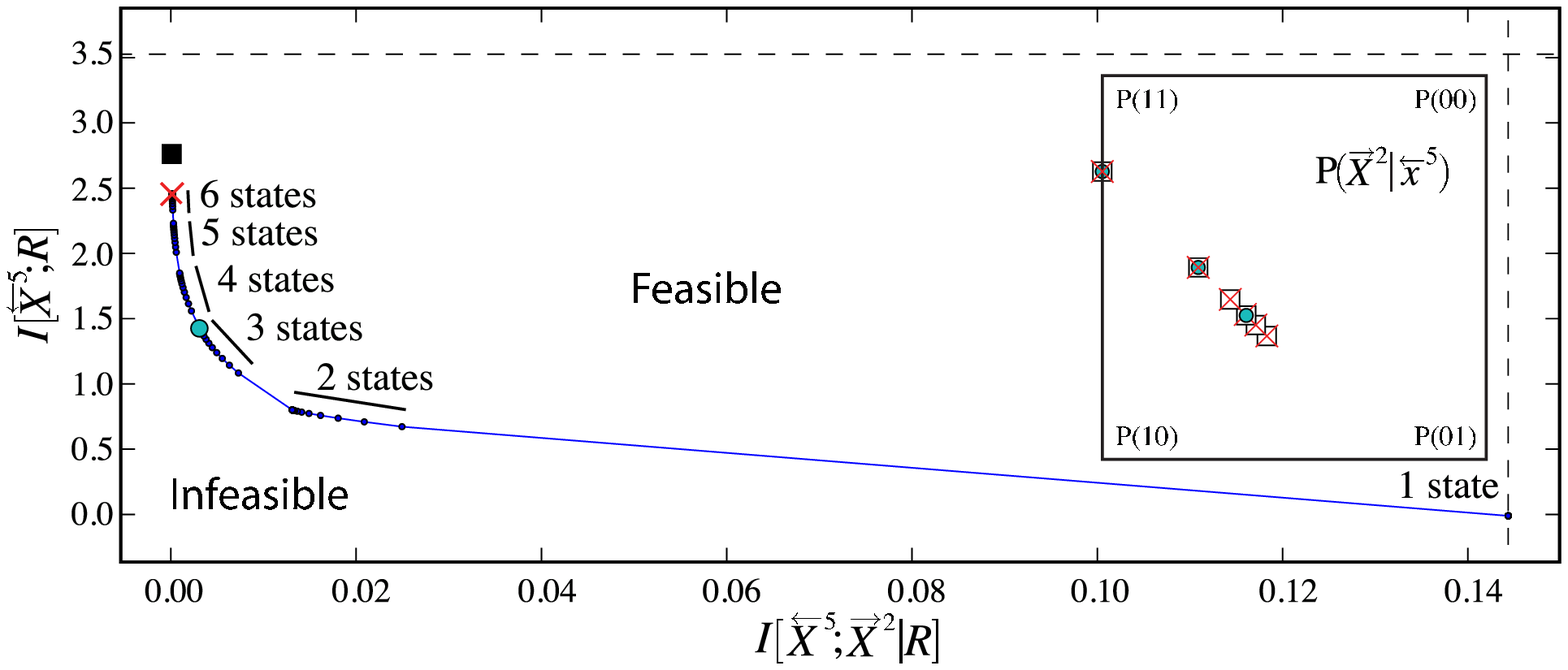}}
\end{center}
\caption{
  Trading structure off against noise using optimal causal inference (OCI):
  Rate-distortion curve for the SNS process, coding rate
  $I[\Past;\AlternateState]$ versus distortion
  $I[\Past;\Future|\AlternateState]$. Dashed lines mark maximum values: past
  entropy $H[\Past^5]$ (horizontal) and excess entropy
  $I[\Past;\Future]$ (vertical). The causal-state limit for infinite
  sequences is shown in the upper left (solid box).
  (Inset) SNS conditional future distributions $\Prob(\Future^{2}|\past^{5})$:
  OCI six-state reconstruction (six crosses), true causal states (six boxes),
  and three-state approximation (three circles). Annealing rate was
  $\alpha = 1.1$.
  }
\label{fig:OCFSNS}
\end{figure*}

\section{Retrieving the causal-state partition}

A key result is that these optimal solutions retrieve the causal-state
partition in the limit $\beta \rightarrow \infty$, which emphasizes prediction
accuracy \cite[detailed proof]{Stil07b}. To see this first note that as
$\beta \rightarrow \infty$, the optimal assignment becomes deterministic
$\Prob_\mathrm{opt}(\alternatestate|\past) \rightarrow \delta_{\alternatestate, \alternatestate^*(\past)}$, 
where the state $\alternatestate^*(\past)$ to which a past is assigned is the
one minimizing energy, Eq. (\ref{OCF_Energy}). Now, that function 
is zero when the future probability conditioned on the state
equals the future probability conditioned on the past. This means
that, in the limit, all pasts with equal conditional future probability distributions will be assigned to the same state with
$\Prob(\Future|\past) = \Prob(\Future|\alternatestate^*(\past))$, for all
those pasts assigned to the state $\alternatestate^*(\past)$. This yields
exactly the causal-state partition given by the equivalence relation that
arises from Eq. (\ref{eps}). 

Hence, one finds in this limit what we have argued is the goal of predictive 
modeling. Moreover, what was otherwise an ad hoc optimization method has been
given a structural grounding in that it captures a process's intrinsic causal
architecture. Recall that the model complexity $\Cmu$ of the causal-state
partition is minimal among the optimal predictors and so not necessarily equal
to the maximum value of the coding rate
$I[\Past;\AlternateState] \leq H[\Past]$. 

\section{Finding approximate causal representations: Causal Compressibility}

While the causal-state partition captures \emph{all} of the predictive information, less complex
models can be constructed if one allows for larger distortion---accepting
less predictive power. For all models in the optimal family, Eqs.
(\ref{OCF_States})-(\ref{OCF_States_2}), the original process is mapped to the
best causal-state \emph{approximation}, at fixed model complexity. And so we
refer to the resulting method as \emph{optimal causal inference} (OCI). Several
examples are studied in \cite{Stil07b}.

The nature of the trade-off embodied in Eq. (\ref{opt}) can be studied by
evaluating the objective function at the optimum for each value of $\beta$.
The shape of the resulting rate-distortion curve characterizes a process's
\emph{causal compressibility} via the interdependence between
$I[\Past;\AlternateState]$ and $I[\Past;\Future|\AlternateState]$. Since the
variation of the objective function in Eq. (\ref{opt}) vanishes at the optimum,
the curve's slope is
$\delta I[\Past;\AlternateState] / \delta D[\Past;\AlternateState] = - \beta$.
For a given process the rate-distortion curve determines what predictability
the best model at a fixed complexity can achieve and, vice versa, how small a
model can be made at fixed predictability. Below the curve lie
\emph{infeasible} causal compression codes; above are \emph{feasible} larger
models that are no more predictive that those directly on the curve. In
short, the rate-distortion curve determines how to \emph{optimally} trade
structure for noise.

As an example, consider the \emph{simple nondeterministic source} (SNS)---a 
hidden Markov process that specifies a binary information source with 
nontrivial statistical structure, including infinite-range correlations and an
infinite number of causal states \footnote{The SNS has causal states
$\causalstate_i, i = 0, 1, \ldots \infty,$
and output-labeled transition matrices whose nonzero entries are
$T^{(0)}_{i0} = \tfrac{1}{2} (1-\tfrac{1}{i+1})$ and
$T^{(1)}_{i,i+1} = \tfrac{1}{2} (1+\tfrac{1}{i+1})$. It produces $\hmu \approx
0.677867$ bits of information per output symbol and stores
$\Cmu \approx 2.71147$ bits of historical information \cite{Crut92c}.}.

The SNS's rate-distortion curve, calculated for pasts of length 5 and futures
of length 2 is shown in Fig. \ref{fig:OCFSNS}. We computed the curve using a 
deterministic annealing scheme following \cite{Rose98}. One starts at a high 
temperature (low $\beta$) and slowly cools the system, waiting for it to
equilibrate---iterating the self-consistent Eqs.
(\ref{OCF_States})-(\ref{OCF_States_2}) until convergence. At that point one
continues by lowering the temperature ($\beta \leftarrow \alpha \beta$)
by a fixed annealing rate $\alpha > 1$ and equilibrating again. During this
procedure, the number of effective states changes. Starting at high
temperatures, all pasts are assigned to states that are all effectively the
same state, as their predictions are equal. States are allowed to split at
each temperature. One observes the proliferation of more and more states 
as the temperature is lowered, until the causal states emerge in the 
zero-temperature limit.

For the SNS the causal states for past and future strings of \emph{finite}
length are recovered by OCI (cross in upper left). For a comparison, there we
also show the \emph{causal-state limit}, which is calculated analytically for
\emph{infinite} pasts and futures (solid box). 

The curve drops rapidly away from the finite causal-state model with six effective states, indicating that there is
little predictive cost in using significantly smaller models with successively 
fewer effective states. The curve then levels out below three states: smaller
models incur a substantial increase in distortion (loss in predictability)
while little is gained in terms of compression. Quantitatively, specifying the best
four-state model (at $I[\Past;\AlternateState] = 1.92$ bits) leads
to 0.5\% distortion, capturing 99.5\% the SNS's excess entropy. The distortion
increases to 2\% for three states (1.43 bits), 9\% for two states
(0.81 bits), and 100\% for a single state (0 bits). Overall, the three-state
model lies near a knee in the rate-distortion curve and this suggests that it
is a good compromise between model complexity and predictability. 

The inset in Fig. \ref{fig:OCFSNS} shows the reconstructed conditional future
distributions for the optimal three-state and six-state models in the
simplex $\Prob(\Future^2 | \past^5)$. The six-state
model (crosses) reconstructs the true causal-state conditional future
distributions (boxes), calculated from analytically known finite-sequence
causal states. The figure illustrates why the three-state model (circles) is
a good compromise: two of the three-state model's conditional future
distributions capture the two more-distinct SNS conditional future
distributions, and its third one summarizes the remaining, less different,
SNS conditional future distributions.
 
\begin{figure}[th]
\begin{center}
\resizebox{!}{1.6in}{\includegraphics{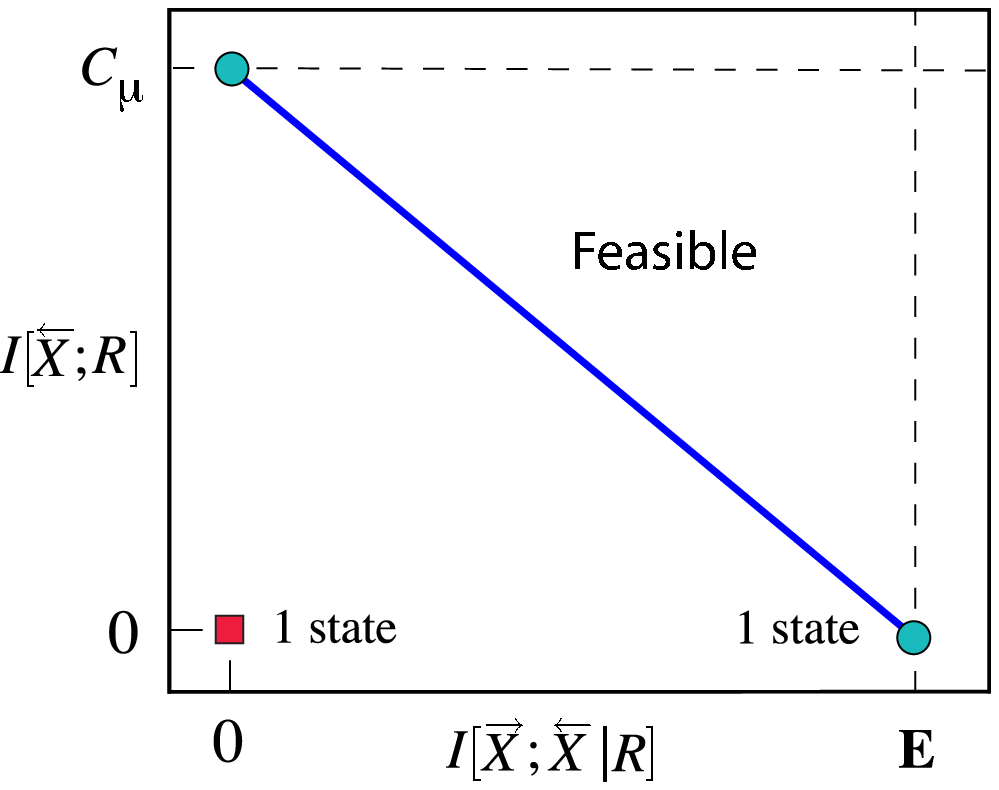}}
\end{center}
\caption{
  Schematic illustration of the causal incompressibility of independent,
  identically distributed processes (square) and predictively reversible
  processes (straight line connecting circles).
  }
\label{fig:NonCompress}
\end{figure}

With its intricate causal structure and nontrivial causal compressibility
properties the SNS process is typical of stochastic processes. Other
frequently studied processes are not, however. Two classes are of particular
interest due to their widespread use. On one extreme of randomness are the
i.i.d. processes alluded to in the introduction, such as the biased
coin---by definition, a completely random and unstructured source.
For all i.i.d. processes the rate-distortion curve collapses to a single
point at $(0,0)$, indicating that they are wholly unpredictable and causally
incompressible. This is easily seen by noting first that for i.i.d. processes 
the excess entropy $I[\Past ;\Future]$ vanishes, since
$\Prob(\future|\past) = \Prob(\future)$. 
Therefore, \mbox{$I[\Past ;\Future | \Partition] \leq I[\Past ;\Future] = 0$}
vanishes, too. Second, the energy function $E(\alternatestate,\past)$
in the optimal assignments,  Eq. (\ref{OCF_Energy}), vanishes, since $\Prob(\future|\alternatestate) =
\left\langle \Prob(\future|\past) \right\rangle_{\Prob(\Past|\alternatestate)}
= \Prob(\future)$. The optimal assignment given by Eq. (\ref{OCF_States}) is
therefore the uniform distribution and
$I[\Past; \AlternateState] \vert_{\Prob_{\rm opt}(\alternatestate|\past)} = 0$.
(See Fig. \ref{fig:NonCompress}.)

At the other extreme are the \emph{predictively reversible} processes for which
\vspace{-0.1in}
\begin{equation}
\Prob(\future|\past) = \delta_{\future,f(\past)},
\label{condition-invert} 
\end{equation}
where $f$ is \emph{invertible}, such as periodic processes. These processes
have a rate-distortion curve that is a straight line, the negative diagonal.
Note that
$\Prob(\future|\alternatestate) = \Prob(f^{-1}(\past)|\alternatestate) = \Prob(\past|\alternatestate)$
and, therefore,
$I[\Past;\Future|\Partition] = I[\Past ;\Future] - I[\Past; \AlternateState]$. 
The variational principle now reads
$\delta (1-\beta) I[\Past; \AlternateState] = 0$, which implies that 
$\beta = 1$. For these processes, the rate-distortion curve is the diagonal
that runs from $[0,\Cmu]$ (causal-state limit) to $[\EE,0]$, where
$\EE = I[\Past;\Future]$ is the excess entropy, due to
Eq. (\ref{condition-invert}) and invertibility. 
(See Fig. \ref{fig:NonCompress}.)

This diagonal rate-distortion curve represents the worst possible case for
causal compression. At each level, specifying the future to one bit higher 
accuracy costs us exactly one bit in model complexity. Processes in this class
are thus not causally compressible. To be causally compressible, a process's
rate-distortion curve must lie below the diagonal. The more concave the
curve, the more causally compressible is the process. An extremely causally
compressible process can be predicted to high accuracy with a model that can
be encoded at a very low model cost. These are the processes that lie between
the extremes of exact predictability and structureless randomness.

These examples show how studying the hierarchy of optimal models, and the
associated rate-distortion curve, allows one to learn about the causal
compressibility of the process at hand, which serves to guide where the
demarcation between structure and noise should lie.

\section{Finite-Sample Fluctuations}

As in statistical mechanics, we assumed so far that the distribution 
$\Prob(\Past,\Future)$ is given. And so, the above results bear on an
intrinsic distinction between structure and noise for a process,
unsullied by statistical sample fluctuations.

However, when one builds a model from \emph{finite} samples, the distributions
must be estimated from the available data and so sample fluctuations must be
taken into account. Intuitively, limited data size sets a bound on how much
we can consider to be structure without overfitting. It turns out that using
\cite{StillBialek2004}, the effects of finite data can be corrected, as we
show in \cite{Stil07b}. This connects the approach taken here to statistical
inference and machine learning, where model complexity control is designed to
avoid overfitting due to finite-sample fluctuations;
cf., e.g, \cite{Wall68a,Akai77a,Riss89a,Vapnik95,MacKay2003}.

\vspace{-0.1in}
\section{Conclusion}

We showed how rate-distortion theory can be employed to find optimal
causal models at varying degrees of abstraction. Starting with the simple
modeling principle of causal shielding, an objective function was constructed
that embodied the trade-off between model complexity and predictability. Since
the variational principle corresponded to a rate-distortion theory known
analysis methods could be employed. Solutions to the objective function were
found using an iterative algorithm, and the rate-distortion curve was
computed using deterministic annealing.

For certain processes we calculated the curve analytically. These and a
numerical example served to demonstrate how its shape reveals a process's
causal compressibility, providing direct guidance for automated model making.
In particular, we showed how a model distinguishes between what it effectively
considers to be underlying structure and what is noise. Practically speaking,
natural processes that have high causal compressibility will admit particularly
parsimonious theories that capture a large fraction of observed behavior.

We pointed out that OCI finds the causal-state partition exactly when the
constraint on model complexity is relaxed. Then we showed how to automatically
build models with varying degrees of abstraction. By focusing on the case
in which limitations due to finite sampling errors are absent, we emphasized
that compact representations, in and of themselves, are critical aids to
scientific understanding. We pointed out, however, that finite data set size
imposes a maximum level of allowable accuracy before overfitting occurs and
that previous results can be used to find that demarcation line as well.
 
\vspace{-0.1in}
\section*{Acknowledgments}

We thank Chris Ellison, on a GAANN fellowship, for programming. The CSC
Network Dynamics Program funded by Intel Corporation supported this work.

\bibliography{ref,chaos,bibthesis}

\begin{thebibliography}{10}

\bibitem{LHC07a}
W.~von Rueden and R.~Mondardini.
\newblock The {Large Hadron Collider (LHC)} data challenge.
\newblock Technical report, IEEE Technical Committee on Scalable Computing,
  2007.
\newblock \url{http://www.ieeetcsc.org/newsletters/2003-01/mondardini.html}.

\bibitem{LSST07a}
Anonymous.
\newblock {LSST} observatory---{Baseline} configuration.
\newblock Technical report, LSST Corporation, Tucson, AZ, 2007.
\newblock \url{http://www.lsst.org/Science/lsst_baseline.shtml}.

\bibitem{NIST06a}
Anonymous.
\newblock {NIST} 2006 machine translation evaluation official results.
\newblock Technical report, National Institute of Standards and Technologies,
  Washington, DC, 2006.

\bibitem{Newm01a}
M.~E.~J. Newman.
\newblock The structure of scientific collaboration networks.
\newblock {\em Proc. Natl. Acad. Sci. USA}, 78(2):404--409, 2001.

\bibitem{Crut88a}
J.~P. Crutchfield and K.~Young.
\newblock Inferring statistical complexity.
\newblock {\em Phys. Rev. Let.}, 63:105--108, 1989.

\bibitem{Crut92c}
J.~P. Crutchfield.
\newblock The calculi of emergence: Computation, dynamics, and induction.
\newblock {\em Physica D}, 75:11--54, 1994.

\bibitem{Crut98d}
J.~P. Crutchfield and C.~R. Shalizi.
\newblock Thermodynamic depth of causal states: {O}bjective complexity via
  minimal representations.
\newblock {\em Phys. Rev. E}, 59(1):275--283, 1999.

\bibitem{Occam}
William~of Ockham.
\newblock {\em Philosophical Writings: {A} Selection, Translated, with an
  Introduction, by {Philotheus} {Boehner}, {O.F.M.}, Late {Professor} of
  {Philosophy}, The {Franciscan Institute}}.
\newblock Bobbs-Merrill, Indianapolis, 1964.
\newblock first pub. various European cities, early 1300s.

\bibitem{Domi99a}
P.~Domingos.
\newblock The role of {Occam's Razor} in knowledge disocvery.
\newblock {\em Data Mining and Knowledge Discovery}, 3:409--425, 1999.

\bibitem{Casd91a}
M.~Casdagli and S.~Eubank, editors.
\newblock {\em Nonlinear Modeling}, SFI Studies in the Sciences of Complexity,
  Reading, Massachusetts, 1992. Addison-Wesley.

\bibitem{Spro03a}
J.~C. Sprott.
\newblock {\em Chaos and Time-Series Analysis}.
\newblock Oxford University Press, Oxford, UK, second edition, 2003.

\bibitem{Kant06a}
H.~Kantz and T.~Schreiber.
\newblock {\em Nonlinear Time Series Analysis}.
\newblock Cambridge University Press, Cambridge, UK, second edition, 2006.

\bibitem{Crut87a}
J.~P. Crutchfield and B.~S. McNamara.
\newblock Equations of motion from a data series.
\newblock {\em Complex Systems}, 1:417 -- 452, 1987.

\bibitem{Cove06a}
T.~M. Cover and J.~A. Thomas.
\newblock {\em Elements of Information Theory}.
\newblock Wiley-Interscience, New York, second edition, 2006.

\bibitem{Crut01a}
J.~P. Crutchfield and D.~P. Feldman.
\newblock Regularities unseen, randomness observed: Levels of entropy
  convergence.
\newblock {\em CHAOS}, 13(1):25--54, 2003.

\bibitem{BialekNemTishby2001}
W.~Bialek, I.~Nemenman, and N.~Tishby.
\newblock {Predictability, Complexity and Learning}.
\newblock {\em Neural Computation}, 13:2409--2463, 2001.

\bibitem{Shannon48}
C.~E. Shannon.
\newblock {A mathematical theory of communication}.
\newblock {\em Bell Sys. Tech. J.}, 27, 1948.
\newblock Reprinted in C. E. Shannon and W. Weaver {\it The Mathematical Theory
  of Communication}, University of Illinois Press, Urbana, 1949.

\bibitem{IBN}
N.~Tishby, F.~Pereira, and W.~Bialek.
\newblock {The information bottleneck method}.
\newblock In B.~Hajek and R.~S. Sreenivas, editors, {\em Proc. 37th Allerton
  Conference}, pages 368--377. University of Illinois, 1999.

\bibitem{Rose98}
K.~Rose.
\newblock {Deterministic Annealing for Clustering, Compression, Classification,
  Regression, and Related Optimization Problems}.
\newblock {\em Proc. IEEE}, 86(11):2210--2239, 1998.

\bibitem{Chec05a}
G.~Chechik, A.~Globerson, N.~Tishby, and Y.~Weiss.
\newblock Information bottleneck for {Gaussian} variables.
\newblock {\em J. Machine Learning Res.}, 6:165--188, 2005.

\bibitem{Creu08a}
F.~Creutzig.
\newblock personal communication, 2008.

\bibitem{Stil07b}
S.~Still, J.~P. Crutchfield, and C.~J. Ellison.
\newblock Optimal causal inference.
\newblock 2007.
\newblock Santa Fe Institute Working Paper 2007-08-024; arxiv.org:0708.1580
  [cs.IT].

\bibitem{StillBialek2004}
S.~Still and W.~Bialek.
\newblock {How many clusters? An information theoretic perspective}.
\newblock {\em Neural Computation}, 16(12):2483--2506, 2004.

\bibitem{Wall68a}
C.~Wallace and D.~Boulton.
\newblock An information measure for classification.
\newblock {\em Comput. J.}, 11:185, 1968.

\bibitem{Akai77a}
H.~Akaike.
\newblock An objective use of {Bayesian} models.
\newblock {\em Ann. Inst. Statist. Math.}, 29A:9, 1977.

\bibitem{Riss89a}
J.~Rissanen.
\newblock {\em Stochastic Complexity in Statistical Inquiry}.
\newblock World Scientific, Singapore, 1989.

\bibitem{Vapnik95}
V.~Vapnik.
\newblock {\em {The Nature of Statistical Learning Theory}}.
\newblock Springer Verlag, New York, 1995.

\bibitem{MacKay2003}
D.~MacKay.
\newblock {\em Information Theory, Inference, and Learning Algorithms}.
\newblock Cambridge University Press, Cambridge, 2003.

\end{thebibliography}

\end{document}